\documentclass[aip,pop,floats,reprint,superscriptaddress]{revtex4-1}
\usepackage{amssymb}
\usepackage{amsmath}
\usepackage{graphicx}

 \def\ds{\displaystyle}
 \def\bc{\begin{center}}          \def\ec{\end{center}}
 \allowdisplaybreaks
\begin{document}
 \title{Response of narrow cylindrical plasmas to dense charged particle beams}
 \author{A.A.Gorn}
 \author{P.V.Tuev}
 \affiliation{Novosibirsk State University, 630090, Novosibirsk, Russia}
 \affiliation{Budker Institute of Nuclear Physics SB RAS, 630090, Novosibirsk, Russia}
 \author{A.V.Petrenko}
 \affiliation{Budker Institute of Nuclear Physics SB RAS, 630090, Novosibirsk, Russia}
 \affiliation{CERN, CH-1211 Geneva 23, Switzerland}
 \author{A.P.Sosedkin}
 \author{K.V.Lotov}
 \affiliation{Novosibirsk State University, 630090, Novosibirsk, Russia}
 \affiliation{Budker Institute of Nuclear Physics SB RAS, 630090, Novosibirsk, Russia}
 \date{\today}
 \begin{abstract}
By combining the linear theory and numerical simulations, we study the response of a radially bounded axisymmetric plasma to relativistic charged particle beams in a wide range of plasma densities. We present analytical expressions for the magnetic field generated in the dense plasma, prove vanishing of the wakefield potential beyond the trajectory of the outermost plasma electron, and follow the wakefield potential change as the plasma density decreases. At high plasma densities, wavefronts of electron density and radial electric field are distorted because of beam charge and current neutralization, while wavefronts of wakefield potential and longitudinal electric field are not. At plasma densities lower than or of the order of beam density, multiple electron flows develop in and outside the plasma, resulting in nonzero wakefield potential around the plasma column.
 \end{abstract}
 \maketitle

\section{Introduction}

Plasma response to relativistic particle beams is a classical problem of plasma physics, actively studied since early 1970s.\cite{PF13-182,PF13-1831,JETP34-93,TP16-1989,PP15-429} The advent of particle beam-driven plasma wakefield acceleration (PWFA)\cite{PRL54-693,Sci.Am.260,PFB5-2363,IEEE-PS24-252} renewed the interest in this problem.\cite{PAcc20-171,PAcc22-81,PF30-252,PoP3-2753,PRE69-046405,PRST-AB7-061302,PRL96-165002,PoP13-056709} PWFA is now pursued as a prospective path to future high-energy accelerators.\cite{NIMA-410-388,NIMA-410-532,PRST-AB5-011001,PoP13-055503,UFN55-965,PPCF56-084013, IPAC14-3791,EPJC76-463,RAST9-63,RAST9-85}
Development of this concept gave impetus to in-depth studies of various special cases, one of which is the response of radially-bounded plasmas to ultra-relativistic particle beams.

Studies of radially-bounded plasmas at setups having a direct relationship to PWFA also started in the 1970s.\cite{JETP39-661,FP2-49} The problem was solved in the linear approximation for uniform plasmas and beams of densities much lower than the plasma density. Later studies focus on effects of radial plasma non-uniformity\cite{PRE60-6210,PoP23-013109}, long-term evolution of nonlinear plasma waves\cite{PRL112-194801}, and beam instabilities\cite{PoP21-056703}.

\begin{figure*}[tb]
\includegraphics[width=\textwidth]{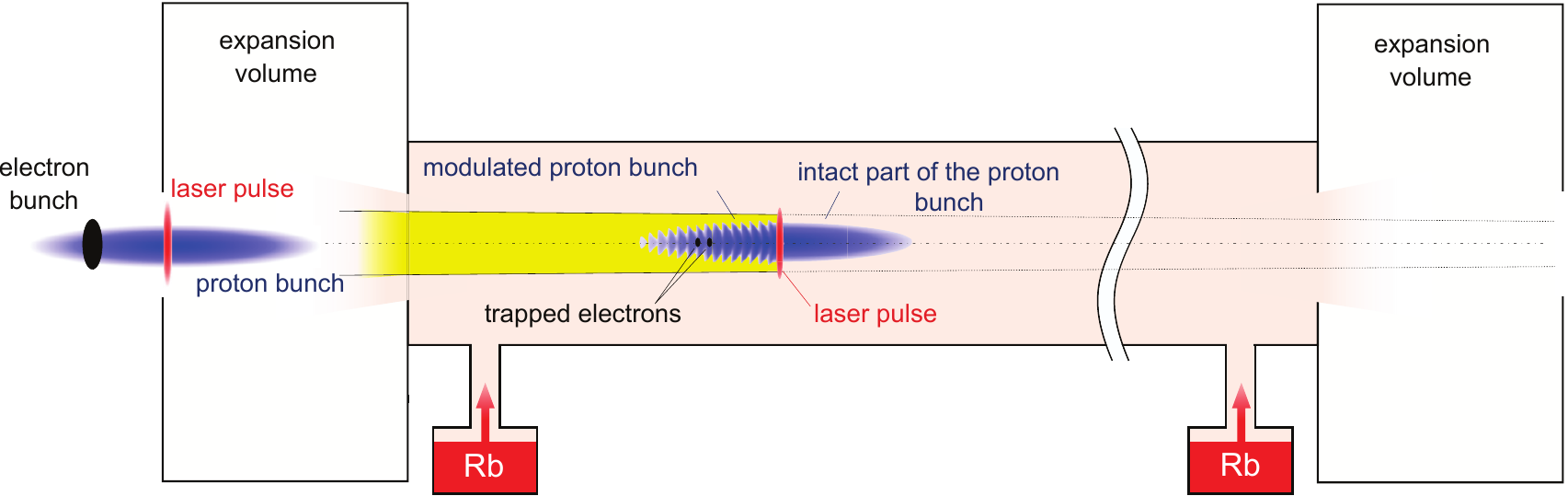}
\caption{Schematic layout of the AWAKE experiment.}\label{fig1-AWAKE}
\end{figure*}

Recently, the experiment AWAKE\cite{PPCF56-084013} at CERN have generated interest in interaction of dense proton beams with low-density plasmas. In AWAKE, three overlapping beams (laser, proton, and electron) propagate through the 10~meter long gas cell filled with the rubidium vapor (Fig.\,\ref{fig1-AWAKE}).\cite{NIMA-829-76,PPCF60-014046} The short laser pulse creates the uniform plasma column with a sharp boundary.\cite{NIMA-740-197} The proton beam self-modulates\cite{PRL104-255003,PoP22-103110} and drives a high-amplitude plasma wave that is witnessed by the electron beam.\cite{PoP21-123116,NIMA-829-3} Since the laser pulse cannot penetrate foils, there are orifices between the gas cell and high-vacuum upstream and downstream beam lines.\cite{NIMA-829-3} The rubidium vapor leaks through the orifices and condenses on cold walls of expansion volumes attached to both ends of the gas cell. The loss of vapor is refilled by two rubidium sources near both orifices so that the vapor flows only near the ends of the cell.\cite{JPDAP-51-025203} The laser pulse ionizes the vapor and creates a radially uniform plasma of an approximately constant radius and the density that gradually reduces away from the orifice. The wakefields excited in the low-density plasma before the orifice by the particle beams are rather weak to disturb the high-energy proton beam, but sufficient for changing trajectories of lower-energy electrons and modifying electron trapping conditions.\cite{PoP21-123116, NIMA-829-3}

In this paper, we consider wakefields driven in radially-bounded low-density plasmas by beams of both charge signs. We follow the plasma response from the linear to strongly nonlinear interaction regime as the plasma density reduces. Beam instabilities possible in plasmas at long interaction times are beyond the scope of this paper. In Sec.\,\ref{s2}, we introduce the model and discuss possible regimes of beam-plasma interaction. In Sec.\,\ref{s3} and Sec.\,\ref{s4} we respectively study linear and nonlinear regimes of the plasma response. In Sec.\,\ref{s5} we discuss how the wakefield in the low-density part of the plasma column modifies electron trapping conditions in the AWAKE experiment. Then in Sec.\,\ref{s6} we summarize the main findings.

\begin{figure}[tb]
\includegraphics[width=\columnwidth]{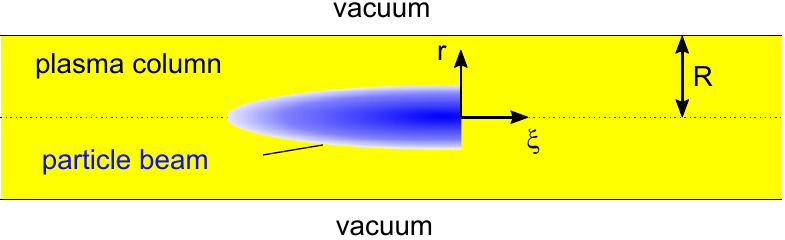}
\caption{Geometry of the problem.}\label{fig2-setup}
\end{figure}

\section{Problem definition and interaction regimes}
\label{s2}

We consider axisymmetric beams and use cylindrical coordinates $(r, \phi, z)$ and the co-moving coordinate $\xi=z-ct$, where $c$ is the speed of light. The plasma column has the radius $R$ and constant density $n_0$ (Fig.\,\ref{fig2-setup}). The plasma is collisionless, and the plasma ions are immobile. The beam density $n_b (r, \xi)$ does not evolve in the co-moving frame. The latter approximation is valid if the beam is ultra-relativistic, and the time scale of beam evolution is much longer than the beam duration. We fix the spotlight on the wakefield potential $\Phi$ that characterizes both focusing and accelerating properties of the plasma wave: the force components acting on an axially moving ultra-relativistic elementary charge $e>0$ are
\begin{equation}\label{e0a}
    e (E_r - B_\phi) = -e \frac{\partial \Phi}{\partial r}, \qquad  e E_z = -e \frac{\partial \Phi}{\partial z},
\end{equation}
where $\vec{E}$ and $\vec{B}$ are electric and magnetic fields. We also focus on the density of plasma electrons $n_e$, as it gives a general idea of plasma response, and discuss other wakefield features as necessary.

\begin{table}[tb]
 \caption{Baseline parameter set for simulations.}\label{t1}
 \bc\begin{tabular}{ll}\hline
  Parameter, notation & Value \\ \hline
  Maximum plasma density, $n_{e0}$ \quad & $7\times 10^{14}\,\text{cm}^{-3}$ \\
  Plasma radius, $R$, & 1.4\,mm \\
  Maximum beam density, $n_{b0}$ \quad & $4\times 10^{12}\,\text{cm}^{-3}$ \\
  Beam total length, $L$ & 30\,cm \\
  Beam radius, $\sigma_r$ & 0.2\,mm \\
  \hline
 \end{tabular}\ec
\end{table}

Some of the considered interaction regimes are intractable analytically. To get insight into their properties, we make numerical simulations with two-dimensional axisymmetric fully kinetic quasistatic code LCODE\cite{PRST-AB6-061301,NIMA-829-350}. Since our study is motivated by AWAKE experiment, we take baseline AWAKE parameters\cite{PoP21-123116} as the reference case (Table~\ref{t1}) and vary the plasma density only. This will limit the variety of interaction regimes to those of known practical importance. Note that in our case the peak beam current is much smaller than $mc^3/e \approx 17$\,kA, where $m$ is the electron mass, and there is the hierarchy of scales
\begin{equation}\label{e2}
    L \gg R \gg \sigma_r.
\end{equation}
For a larger beam current and different ratio $\sigma_r/R$, the interaction regimes could be different.

The particular beam shape is
\begin{equation}\label{e1}
    n_b (r, \xi) = \begin{cases}
    n_{b0} e^{-r^2/2 \sigma_r^2} \bigl(1 + \cos(\pi\xi/L)\bigr)/2, & -L<\xi<0, \\
    0, & \text{otherwise}.
    \end{cases}
\end{equation}
It is rather convenient for basic studies because it has both slowly varying (long tail) and rapidly changing (hard leading edge) parts, so the study can inform of the plasma response on beams of different timescales. While our focus is on positively charged beams, we also consider electron beams wherever comparison of the two cases is helpful.

\begin{table}[tb]
 \caption{Boundaries between the interaction regimes and effects responsible for this boundaries.}\label{t2}
 \bc\begin{tabular}{llrc}\hline
  Equality & Effect & Plasma density \ & $n_0/n_{b0}$ \\ \hline
  $k_p R = 1$ & plasma boundary & $1.5 \times 10^{13}\text{cm}^{-3}$ & 3.6 \\
  $n_{b0} = n_0$ & plasma nonlinearity & $4\times 10^{12}\,\text{cm}^{-3}$ & 1.0 \\
  $n_0 R^2 = 2 n_{b0} \sigma_r^2$ & plasma self-fields & $4\times 10^{10}\,\text{cm}^{-3}$ & 0.01 \\
  \hline
 \end{tabular}\ec
\end{table}

For the selected relation of scales, we can distinguish four regimes of plasma response (Table~\ref{t2}). The first regime corresponds to high plasma densities, where the plasma radius $R$ is much larger than the plasma skin-depth $k_p^{-1} = c/\omega_p$, where $\omega_p = \sqrt{4 \pi n_0 e^2/m}$ is the plasma frequency. In this regime, there is no difference between unbounded and radially bounded plasmas.

In the second regime, effects of the plasma boundary are important ($k_p R \lesssim 1$), but still $n_0 \gg n_{b0}$, and nonlinear effects are weak. The first two regimes allow for a unified analytical description (Sec.\,\ref{s2}).

In the third regime, $n_0 \lesssim n_{b0}$, and the plasma response is strongly nonlinear (Sec.\,\ref{s3}). Still, the plasma column contains enough plasma electrons to neutralize the beam charge and current.

The fourth regime corresponds to very low plasma densities. In this regime (also described in Sec.\,\ref{s3}), the beam linear charge exceeds that of the plasma column, $n_b \sigma_r^2 > n_0 R^2$, and plasma fields has a negligible effect on the motion of plasma electrons.

The transition between the regimes is smooth and the equalities presented in Table~\ref{t2} show the transition borders only approximately.

\section{Linear plasma response}
\label{s3}

The expressions for wakefields induced in the radially-bounded uniform plasma by a low-density particle beam have an easy-to-use form\cite{PoP21-056703,notebook} if the beam density is separable (as in our case),
\begin{equation}\label{e4a}
    n_b(r,\xi) = n_{b0} f(r) g(\xi).
\end{equation}
Then the potential is also separable
\begin{align}
    \label{e6a} \Phi (r,\xi) &= \begin{cases}
\ds q\frac{mc^2 n_{b0}}{e n_0} F(r) G(\xi), & r<R, \\
0, & r>R,
\end{cases} \\
    \label{e7b} G(\xi) &= k_p \int_\xi^\infty d\xi' \sin \bigl( k_p (\xi'-\xi) \bigr) g(\xi'), \\
    \nonumber F(r) &= k_p^2 \int_0^R \left[\frac{K_0(k_p R)}{I_0(k_p R)}I_0(k_p r_>) - K_0(k_p r_>)\right] \\
    \label{e7a} &\times I_0(k_p r_<) f(r') r'dr',
\end{align}
where
\begin{equation}\label{e7c}
    r_< = \min (r, r'), \qquad r_> = \max (r, r'),
\end{equation}
$q=\pm 1$ is the beam charge sign, and $I_0$ and $K_0$ are modified Bessel functions. Note the same longitudinal periodicity of the potential at all radial positions and no boundary effect on the oscillation frequency in the near-boundary regions.

\begin{figure}[tb]
  \includegraphics[width=236bp]{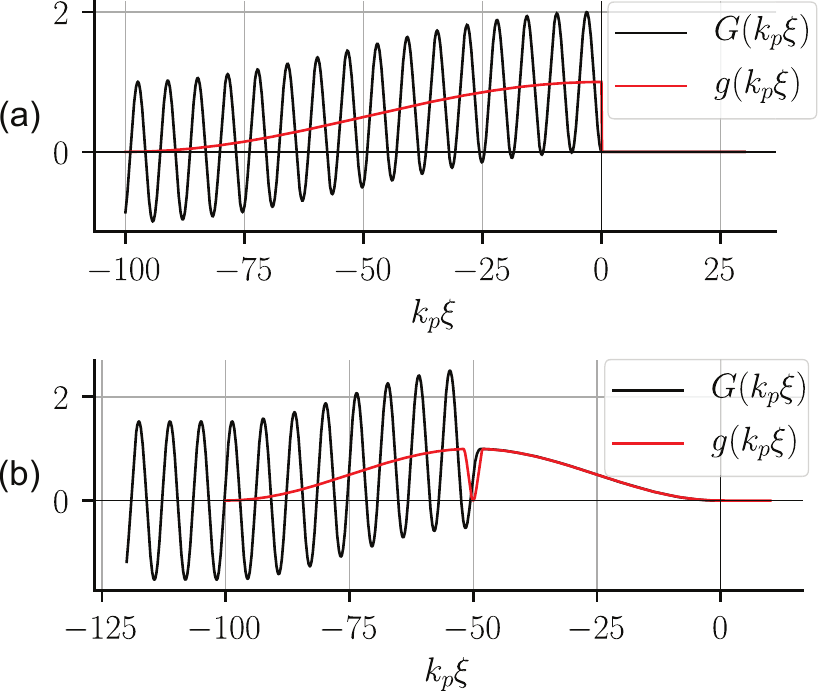}
  \caption{Longitudinal functions $g(\xi)$ (beam shape) and $G(\xi)$ (wakefield potential) for beams with (a) sharp leading edge and (b) localized short-scale fragment.}\label{fig-G}
\end{figure}

Properties of the longitudinal function \eqref{e7b} are best seen after integrating by parts
\begin{multline}\label{e8}
    G(\xi) = g(\xi) + \sin (k_p \xi) \int_\xi^\infty \sin (k_p \xi') \frac{d g (\xi')}{d \xi'} \, d \xi' \\
    + \cos (k_p \xi) \int_\xi^\infty \cos (k_p \xi') \frac{d g (\xi')}{d \xi'} \, d \xi'.
\end{multline}
The first term follows the beam density profile and can be slowly-varying. The second and third terms oscillate with the plasma frequency, and their total amplitude is proportional to the Fourier component of the derivative $d g/d \xi$ at the this frequency. If the beam has a sharp leading edge [Fig.\,\ref{fig-G}(a)], then the amplitude of the oscillating component always equals $g(\xi)$ at the edge location. In the general case, a localized short-scale fragment can initiate oscillations of an arbitrary amplitude [Fig.\,\ref{fig-G}(b)].

\begin{figure}[tb]
\includegraphics[width=228bp]{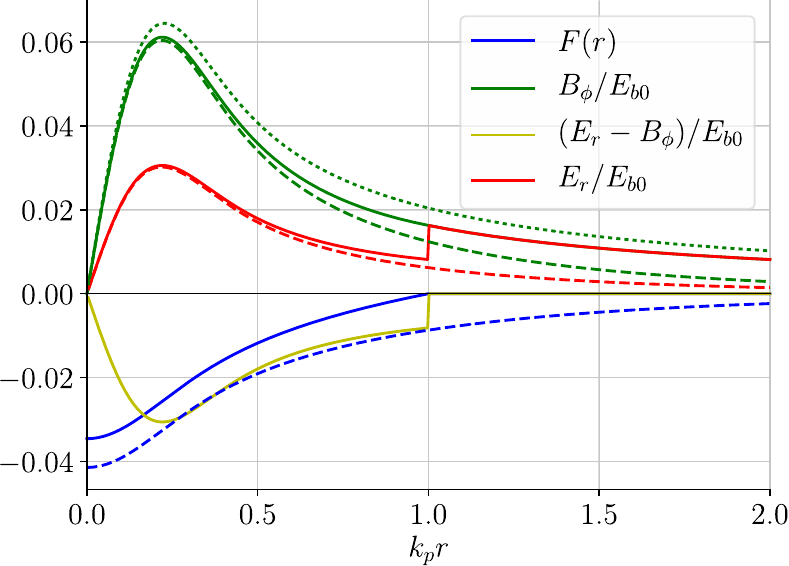}
\caption{Radial dependences of the wakefield potential term $F(r)$, radial force $E_r - B_\phi$, and fields $E_r$ and $B_\phi$ for $k_p R = 1$, $G(\xi) = 0.5$, $g(\xi) = 1$. Dashed lines show the corresponding dependences for the unbounded plasma. The dotted line is the vacuum magnetic field of the beam.}\label{fig-sum}
\end{figure}

The oscillating part of the wakefield appears due to Langmuir waves, which are potential and produce no magnetic field. Accordingly, the expression for the magnetic field $B_\phi$ contains no oscillations at the plasma frequency
\begin{multline}
    \label{e9} B_\phi (r,\xi) = -qE_{b0} k_p g(\xi)\int_0^R dr' r' \frac{d f(r')}{dr'} \\
 \times \begin{cases}
 \left[\alpha I_1(k_p r_>) + K_1(k_p r_>)\right] I_1(k_p r_<), & r<R, \\
 I_1(k_p r') \left(\alpha I_1(k_p R) + K_1(k_p R)\right) R / r, & r>R,
\end{cases}
\end{multline}
where
\begin{equation}
\alpha = \frac{K_0(k_p R)}{I_0(k_p R)},
\end{equation}
and
\begin{equation}\label{e12a}
    E_{b0} = \frac{mc\omega_p n_{b0}}{e n_0}
\end{equation}
is a convenient field unit for our problem. We obtained the formula \eqref{e9} similarly to the infinite plasma case\cite{PP15-429}, but with two additional interface conditions for continuity of $B_\phi$ and $\partial B_\phi / \partial r$ at $r=R$. The expression \eqref{e9} relates to the ultra-relativistic beam case and, therefore, differs from that of Ref.~\onlinecite{JETP39-661}, which corresponds to moderately relativistic beams.

Unlike the wakefield potential, the magnetic field \eqref{e9} does not vanish outside the plasma. Consequently, the radial electric field in the outer region equals the magnetic field for any beam shape and radius (Fig.\,\ref{fig-sum}). Thus, the plasma fields can be conceived as composed of two parts. One part is the plasma wave excited by longitudinal beam non-uniformities. Its frequency equals the plasma frequency, and its field is purely electric and does not extend outside the plasma column. The other part is incompletely neutralized electric and magnetic self-fields of the beam, which may have different radial dependence inside the plasma, but are identical outside. Both parts have the same radial dependence of the wakefield potential. The surface wave\cite{JETP39-661,PoP10-4563} is not excited in our case, as its phase velocity is smaller than the beam (light) velocity.

\begin{figure}[tb]
\includegraphics[width=\columnwidth]{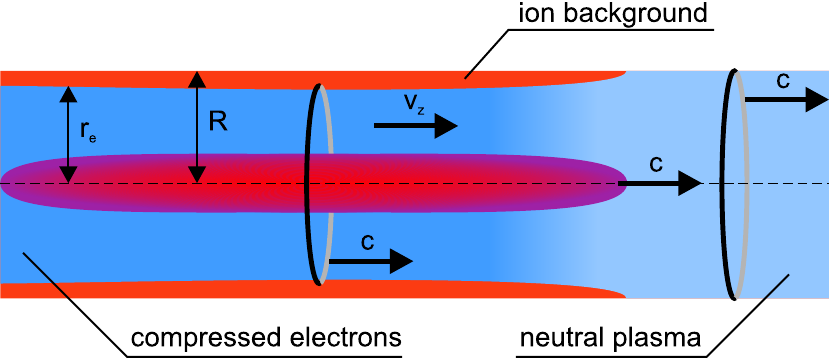}
\caption{ Illustration to the derivation of field equality outside the plasma.}\label{fig-elflux}
\end{figure}

The equality of $E_r$ and $B_\phi$ outside the plasma comes from the electron flux conservation in the co-moving frame (Fig.\,\ref{fig-elflux}) that necessarily takes place in the context of the quasistatic approximation. The number of electrons passing through black circles in Fig.\,\ref{fig-elflux} is the same and equals
\begin{equation}\label{e10}
  \int_0^{r_e} n_e (c - v_z) \, 2 \pi r \, dr = \int_0^R n_0 c \, 2 \pi r \, dr,
\end{equation}
where $v_z (r,\xi)$ and $n_e(r,\xi)$ are longitudinal velocity and density of plasma electrons, $n_0$ is the unperturbed electron density equal to the ion density $n_i$, and $r_e$ is the radius of the outermost electron. Since the beam current $j_{bz} = e n_b c$, from Maxwell and Poisson equations we have
\begin{equation}\label{e11}
    \frac{\partial}{\partial r} r (E_r - B_\phi) = 4 \pi e r \left( n_i - n_e + n_e \frac{v_z}{c} \right),
\end{equation}
which, after integrating and using \eqref{e10}, gives $E_r (r) = B_\phi (r)$, if $r > r_e$ and $r > R$. This is also valid for the nonlinear plasma response and proves the equality of $E_r$ and $B_\phi$ beyond the trajectory of the outermost electron.

\begin{figure}[tb]
\includegraphics[width=\columnwidth]{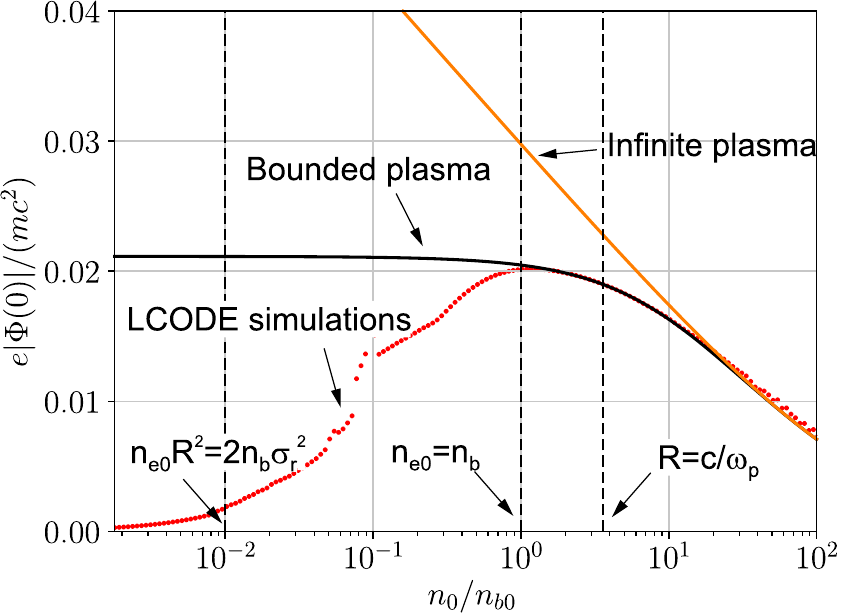}
\caption{ Plasma density dependence of the on-axis potential calculated in the linear approximation for bounded and unbounded plasmas (solid lines) and simulated for the bounded plasma (dots). Vertical lines show the boundaries between the interaction regimes from Table~\ref{t2}.}\label{fig-evolphi}
\end{figure}

Wakefield strength scales differently with the decrease of the plasma density in bounded and unbounded plasmas. The amplitude of the longitudinal function \eqref{e8} does not depend on the plasma density for our beam, so the difference comes from the radial function $F(r)$. In the unbounded plasma, the low-density limit corresponds to $k_p \sigma_r \ll 1$, for which\cite{PoP12-063101}
\begin{equation}\label{e12}
    F(0) \approx k_p^2 \sigma_r^2 [0.05797 - \ln (k_p \sigma_r)],
\end{equation}
and the potential amplitude grows in absolute value, as the density decreases (Fig.\,\ref{fig-evolphi}):
\begin{equation}\label{e13}
    \Phi(0) \propto 0.05797 - \ln (k_p \sigma_r).
\end{equation}
Consequently, the lower the plasma density, the larger emittance the beam must have to stay in equilibrium with the wakefield in the unbounded plasma\cite{PoP24-023119}. For the same reason, particles side-injected into the wakefield\cite{NIMA-829-3,JPP78-455} in a low-density plasma gain a larger transverse momentum than in a high-density plasma. The longitudinal electric field $E_z$, however, is smaller at low densities, as decrease of the derivative $\partial / \partial z \approx k_p \propto \sqrt{n_0}$ in \eqref{e0a} prevails over the slow logarithmic growth \eqref{e13}.

In the bounded plasma, the first term in square brackets in Eq.\,\eqref{e7a} dominates at $k_p R \ll 1$, and the scaling at low plasma densities is
\begin{equation}\label{e14}
    F(0) \approx -\frac{k_p^2 \sigma_r^2}{2} \left(\ln{\frac{R^2}{2 \sigma_r^2}} + \Gamma(0, R^2/(2 \sigma_r^2)) + \gamma \right),
\end{equation}
where
\begin{equation}\label{e14a}
 \Gamma (0, \beta) = \int_\beta^\infty t^{-1} e^{-t} dt, \qquad \gamma \approx 0.577215,
\end{equation}
so the potential well depth tends to a constant (Fig.\,\ref{fig-evolphi}). At even lower densities, for which the linear theory is not applicable and simulations are needed, the potential well gradually disappears (Fig.\,\ref{fig-evolphi}). The linear theory thus gives a correct value of the potential well depth up to the onset of nonlinear effects at $n_0 \sim n_{b0}$.

\section{Nonlinear plasma response}
\label{s4}

It is commonly believed that the linear theory of plasma response to the beam is fully applicable if the plasma density is much higher than the beam density,
\begin{equation}\label{e15}
n_0 \gg n_b.
\end{equation}
Evaluation of all neglected nonlinear terms\cite{PP15-429,PoP3-2753} formally gives a stronger limitation
\begin{equation}\label{e16}\
n_0 \gg n_b k_p^2 L^2,
\end{equation}
but weakly restricts the validity of the linear theory results, as account of the nonlinear terms does not considerably change the plasma response.

\begin{figure}[tb]
\includegraphics[width=\columnwidth]{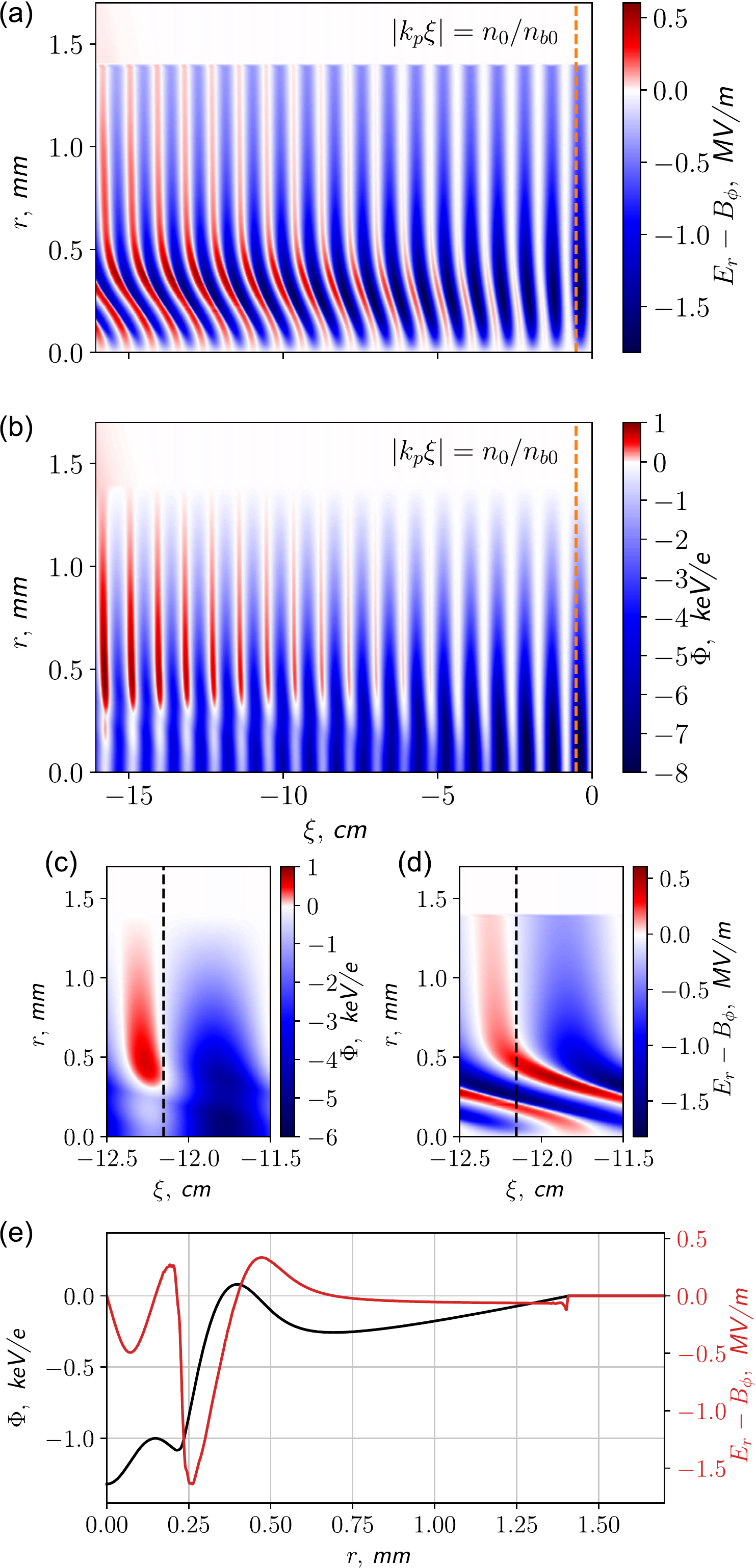}
\caption{ Maps of the radial force $E_r -B_\phi$ and wakefield potential $\Phi$ in wide (a), (b) and zoomed in (c), (d) areas and their radial slices (e) at $\xi = -12.15\,\text{cm}$ (black dashed line) for the plasma density $n_0 \approx 3.6 n_{b0}$ (at which $k_p R = 1$).}\label{fig-drift}
\end{figure}

However, if the beam has two different scales (as in our case), the applicability condition for the linear theory is much stronger,
\begin{equation}\label{e17}
n_0 \gg n_b k_p L,
\end{equation}
where $L$ is the larger scale. The limitation comes from changing the local plasma frequency due to beam charge neutralization and from the drift of plasma electrons, neutralizing the beam current.
Plasma electron density perturbation $\delta n$ and longitudinal velocity $v_z$ are
\begin{equation} \label{e17a}
\delta n = qn_{b0} f(r)G(\xi),  \qquad v_z = qc\frac{n_{b0}}{n_0}F(r)G(\xi).
\end{equation}
These quantities, averaged over the plasma wave period, determine the plasma frequency shift. For smooth drivers [$|dg(\xi)/d \xi| \ll k_p$], the averaging takes a simple form
\begin{equation} \label{e17c}
\langle G(\xi) \rangle = g(\xi).
\end{equation}
Therefore, the average density perturbation copies the shape of the driver beam while the speed of plasma electrons \eqref{e17a} copies the shape of the wakefield potential.

If the ratio $n_b/n_0$ is small, the plasma frequency changes by $\delta \omega_p \sim q \omega_p n_b/(2n_0)$ in the beam area. Plasma electrons move with the average velocity $v_z \sim c n_b / n_0$ (in the case of local neutralization for $k_p \sigma_r \gg 1$) or less (if the plasma current flows in a wider area for $k_p \sigma_r \lesssim 1$). The electron motion causes the Doppler shift of the oscillation frequency by about $q\omega_p n_b/n_0$ or less. Two effects add together and result in deformation of wavefronts, which accumulates towards the beam tail [Fig.\,\ref{fig-drift}(a)]. The sense of curvature depends on the beam charge sign. The limitation \eqref{e17} comes from the requirement of a small accumulated phase shift. The analogue to this is a limitation on the distance $|\xi|$ from the beam part that generates the plasma wave:
\begin{equation}\label{e18}
|\xi| \ll k_p^{-1} n_0 / n_{b0}.
\end{equation}
At these distances, the linear theory gives correct fields, velocities, and electron densities (orange dashed lines in Fig.\,\ref{fig-drift}).

Surprisingly, the distortion of wavefronts does not result in distortion of the wakefield potential pattern [Fig.\,\ref{fig-drift}(b)]. The potential $\Phi$ and its longitudinal derivative $E_z$ oscillate exactly with the plasma frequency, while patterns of the radial force $E_r - B_\phi$ and plasma electron density $n_e$ are distorted and have a period, which is shorter or longer than the plasma period, depending on the driver charge sign. This unusual feature appears due to low-amplitude, short-scale radial rippling of the potential [Fig.\,\ref{fig-drift}(c)] and violation of separability \eqref{e6a}.

\begin{figure}[tb]
\includegraphics[width=211bp]{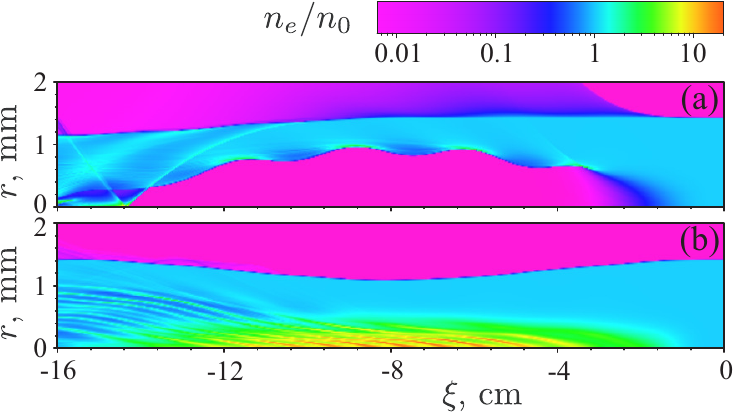}
\caption{ Maps of the plasma electron density $n_e$ for smooth electron (a) and proton (b) beams of the shape \eqref{e19}. The initial plasma density is $n_0= 0.1 n_{b0}.$}\label{fig-channel}
\end{figure}

At densities $n_0 \lesssim n_{b0}$, the plasma response is strongly nonlinear. We started studying it from the case of long smooth beams of both charge signs with the density distribution
\begin{equation}\label{e19}
    n_b (r, \xi) = \begin{cases}
    n_{b0} e^{-r^2/2 \sigma_r^2} \bigl(1 - \cos(2\pi\xi/L)\bigr)/2, & -L < \xi < 0, \\
    0, & \text{otherwise}.
    \end{cases}
\end{equation}
For electron beams, the main difference from the unbounded plasma case is that electrons initially located in outer layers leave the plasma, as the driver current increases, and carries away the excessive negative charge [Fig.\,\ref{fig-channel}(a)]. When the driver current later decreases, these electrons cannot quickly return, so the plasma acquires a positive charge and generates a radial electric field. The column of plasma electrons shrinks in radius to keep the charge balance inside, so the total positive charge of the plasma is that of bare ions in the outer layer. Small plasma-frequency oscillations of the radial electric field, which are always present in this system, cause some boundary electrons to gain a large inward momentum and form a multiple flow inside the plasma. Formation of the ion channel (or bubble) near the axis does not differ from the unbounded plasma case.\cite{PRE69-046405} For the proton driver, no electrons escape the plasma [Fig.\,\ref{fig-channel}(b)], and the excessive positive charge of the beam transfers to an ``ion tube'' in the outer region of the plasma column.

\begin{figure}[tb]
\includegraphics[width=212bp]{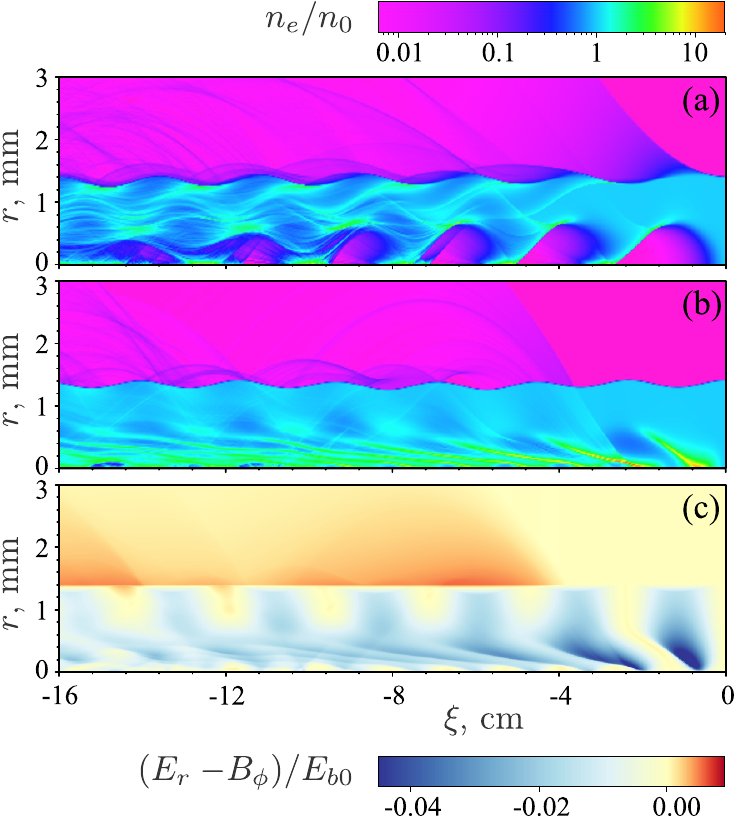}
\caption{ Maps of the plasma electron density $n_e$ for electron (a) and proton (b) beams of the shape \eqref{e1}. Map of the focusing force $E_r-B_\phi$ for the proton beam case (c). The initial plasma density is $n_0= 0.5 n_{b0}.$}\label{fig-outer}
\end{figure}

If the beam efficiently generates both low-frequency and plasma-frequency perturbations, the number of electrons escaping the plasma is even higher. For the electron beam, nonlinear oscillations of the plasma electron density near the axis cause oscillations of the electron boundary [Fig.\,\ref{fig-outer}(a)]. During each oscillation period, groups of electrons separate from the boundary and either leave the plasma column, or propagate towards the axis forming a multiple flow. The escaping electrons appear also as a result of wave breaking in the near-axis region. For the proton beam, the escaping electrons appear from the wave breaking only [Fig.\,\ref{fig-outer}(b)].

The electrons that escape the plasma column carry a non-zero wakefield potential to the region of their reach and, therefore, make this region defocusing for proton beams and focusing for electron beams [Fig.\,\ref{fig-outer}(c)]. The property of no radial force exerted on ultra-relativistic beams outside the narrow plasma thus disappears at low plasma densities.

\begin{figure}[tb]
\includegraphics[width=\columnwidth]{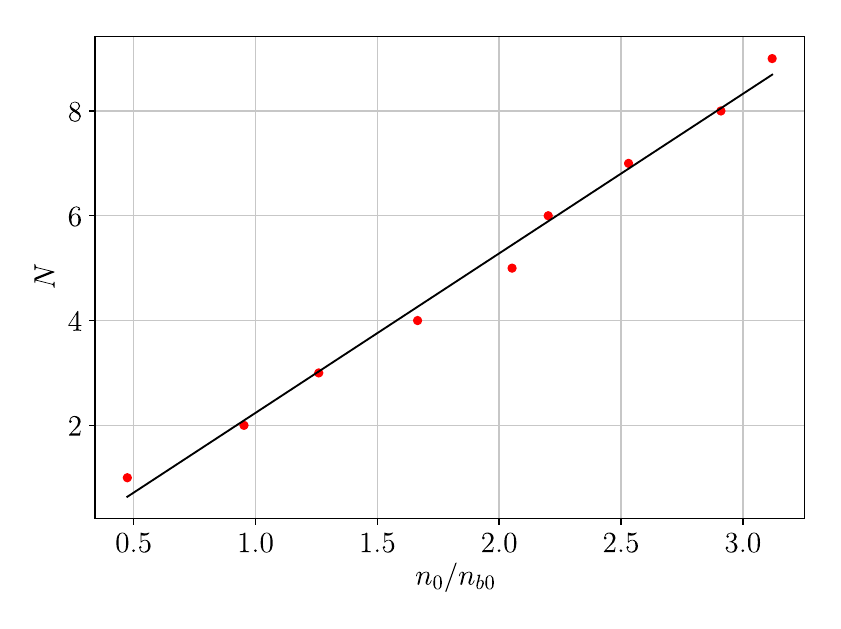}
\caption{Simulated plasma density dependence of the wakefield period number N where the wave first breaks (points). The solid line helps to see the linear scaling.}\label{fig-wbreak}
\end{figure}

Appearance of escaping electrons is directly related to distortion of wavefronts discussed earlier. As the phase difference of electron oscillations at different radii reaches some critical value, electron trajectories cross: inner electrons become outer and vice versa. The escaping electrons are those initially located at smaller radii. After the trajectories cross, these electrons experience the radial expelling force from the increased negative charge inside and, therefore, escape the plasma rather than continue oscillating around some radial position. The lower the plasma density, the stronger the distortion of wavefronts, the sooner the wavebreaking occurs (Fig.\,\ref{fig-wbreak}). Note also that the oscillating component of the radial force almost disappears at radii of wavebreaking [Fig.\,\ref{fig-outer}(c)], only the slowly varying component remains.

\begin{figure}[tb]
\includegraphics[width=212bp]{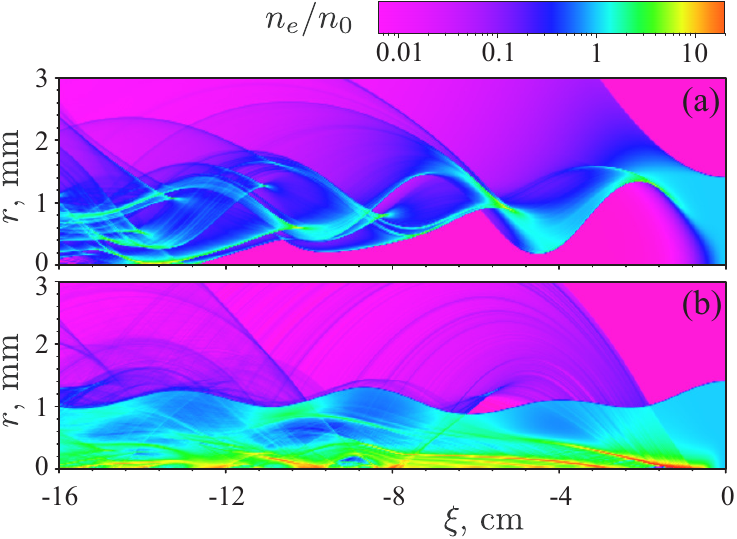}
\caption{ Maps of the plasma electron density $n_e$ for electron (a) and proton (b) beams and initial plasma density $n_0= 0.14 n_{b0}$.}\label{fig-waves}
\end{figure}

For plasma densities satisfying conditions $n_{b0} \gg n_0 \gg n_{b0} \sigma_r^2/R^2$, two different oscillation scales are visible at density maps (Fig.\,\ref{fig-waves}). In parallel with plasma-frequency oscillations, there appears radial oscillations of electrons ejected out of the plasma. The time scale of the radial oscillations depends on the linear charge of the plasma column and on the beam current. Two oscillation types do not continuously evolve into another, thus forming a chaos-like plasma response, if both are present and strong.

\begin{figure}[tb]
\includegraphics[width=212bp]{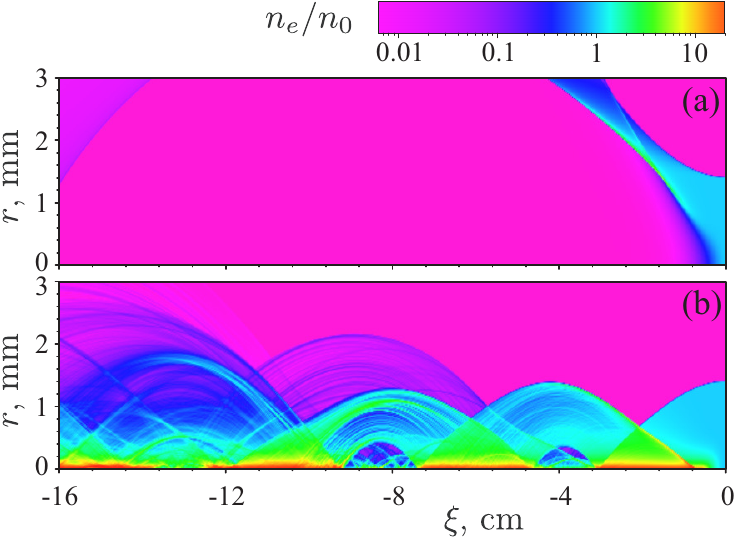}
\caption{ Maps of the plasma electron density $n_e$ for electron (a) and proton (b) beams and initial plasma density $n_0= 0.05 n_{b0}$.}\label{fig-chaos}
\end{figure}

At very low plasma densities, the behavior of plasma electrons and the wakefields are determined by the beam fields (Fig.\,\ref{fig-chaos}). For electron beams, all plasma electrons are ejected out of the plasma and return back well after the beam transit. For proton drivers, the plasma electrons oscillate around the beam. The interaction regime changes to this ``low density'' mode well before the linear charge of plasma electrons equals the beam linear density, as is illustrated in Fig.\,\ref{fig-chaos}.

\section{Problem of electron injection}
\label{s5}

Quantitative measures of the plasma response depend on particular beam and plasma parameters. For this reason, we have discussed mostly qualitative features in the previous section. Here we quantitatively study the effect of smooth density transition at the beginning of the plasma section on propagation of the witness electron beam in the AWAKE experiment. We take the longitudinal plasma density profile\cite{NIMA-829-3, JPDAP-51-025203}
\begin{equation}\label{e20}
  n_0 = \frac{n_{e0}}{2} \left( 1 - \frac{\delta z / D}{\sqrt{(\delta z / D)^2+0.25}} \right),
\end{equation}
where $\delta z$ is the distance to the orifice that separates the plasma cell and expansion volume at $z=0$, and $D=1$\,cm is the orifice diameter.

\begin{figure}[tb]
\includegraphics[width=\columnwidth]{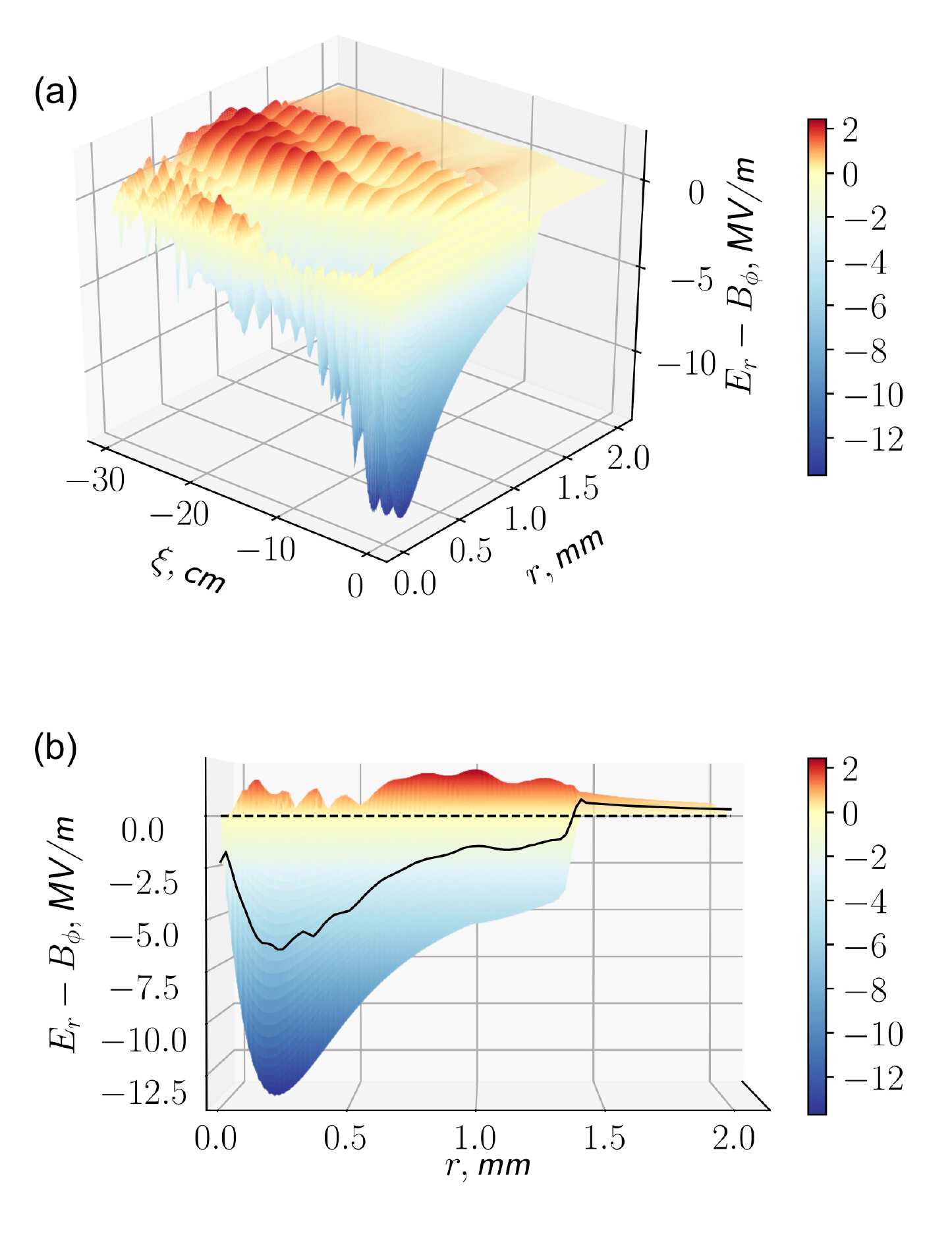}
\caption{ The focusing force $E_r - B_\phi$ (a)  and the ``integral'' radial force $F_\text{r,int} (r)$ (b) for the proton beam and plasma of the density $n_0= n_{b0}$. }\label{fig-outline}
\end{figure}

The optimum injection parameters\cite{NIMA-829-3} were obtained from simulations of electron beam propagation through the whole plasma section, including the density transition areas. In this Section, we present a less precise, but more intuitive explanation why the optimum is like that. For this, we reduce full maps of the radial forces available for each plasma density value [Fig.\,\ref{fig-outline}(a)] to simple radial dependencies $F_\text{r,int} (r)$ showing the integral effect of the radial force [Fig.\,\ref{fig-outline}(b)]. The ``integral'' radial force $F_\text{r,int} (r)$ is the half-sum of the maximum and minimum forces at the given radius. If the wakefield oscillates as a function of $\xi$, as is typical for $r < R$,  then the `integral'' force presents the average slow-varying force component. If the force has a definite sign, as for $r > R$, then the `integral'' force is half of the maximum force at this radius.

\begin{figure}[tb]
\includegraphics[width=\columnwidth]{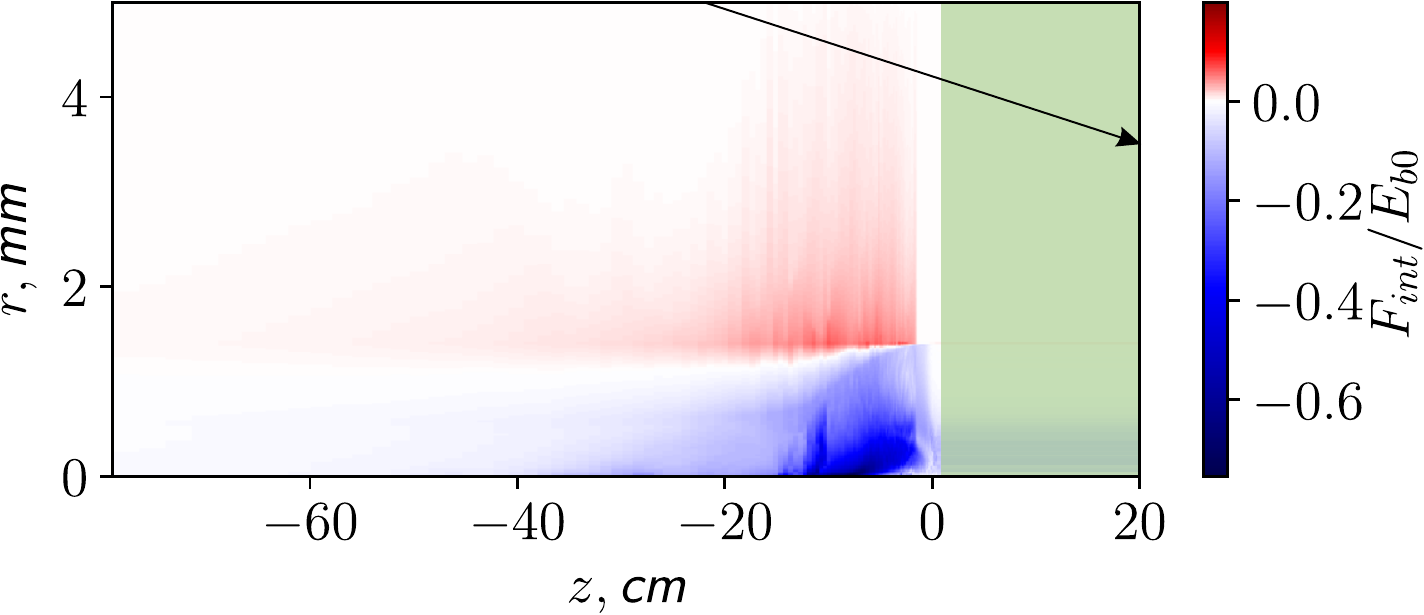}
\caption{ The ``integral'' radial force $F_\text{r,int} (r)$ in the density transition area near the orifice. The arrow shows the optimum electron trajectory for the best injection efficiency (from Ref.\,\onlinecite{NIMA-829-3}). The green shading shows the location of the constant density area, in which electrons can be trapped by the stationary-phase wakefield.}\label{fig-map}
\end{figure}

The map of the ``integral'' radial force (Fig.\,\ref{fig-map}) shows that the plasma column always defocuses axially propagating electrons. The typical defocusing force is several MeV/m, which is sufficient to deflect a 16\,MeV electron beam (the AWAKE design energy) by 1.4\,mm (plasma radius) in several centimeters of propagation. Thus, no collinearly injected electrons can cross blue areas of Fig.\,\ref{fig-map} and get trapped by the driver wakefield in areas of the constant density inside the plasma cell.

\section{Summary}
\label{s6}

We combined the linear analytical theory and numerical simulations to study the response of a radially bounded plasma to highly relativistic charged particle beams in a wide range of plasma densities.

We discovered that the wakefield potential vanishes outside the plasma column. This result is valid for any plasma density, as it is a direct consequence of the axial symmetry and charge conservation. For a strongly nonlinear plasma response, some electrons may leave the plasma and carry away a nonzero potential. If so, the potential vanishes beyond the outermost electron trajectory.

The wakefield potential is strongest in absolute value at plasma densities close to the peak beam density ($n_0 \sim n_b$). At higher plasma densities, the wakefield is tractable analytically and is weaker for the radially bounded case, as compared with the infinite plasma, if the plasma skin depth is longer than or of the order of the plasma radius. At lower plasma densities, analytical expressions overestimate the wakefield amplitude that falls to zero as the density decreases.

For long low-density beams ($n_b \ll n_0$), the nonlinearity of the plasma response manifests itself as wavefront distortion caused by compensation of beam charge and current in the plasma. This imposes strong limitations \eqref{e17} or \eqref{e18} on the applicability of the linear theory. Patterns of the wakefield potential and the longitudinal electric field, however, are not distorted and keep the unperturbed plasma period.

Even at low plasma densities ($n_0 \lesssim n_b$), the plasma maintains average quasi-neutrality in most of its volume. In the case of electron beams, this is achieved by pushing a certain number of plasma electrons out of the plasma column. A positively charged beam pulls all electrons from the near-boundary region and leaves a ``tube'' of bare ions there. Plasma oscillations initiated by the beam, if any, produce electron jets that form multiple flows inside and outside the plasma column. The jets originate either from the plasma boundary or from wavebreaking regions near the axis, and their origin is locked to certain oscillation phases. Timescales of jet dynamics differ from the period of plasma oscillations, so the multiple flow in the presence of strong jets looks chaos-like. The plasma response at very low plasma densities is expectedly fully determined by the beam fields.

The wakefield created by the beam in the plasma is, in average, focusing for this beam and for witness particles of alike charge and is defocusing for particles of the opposite charge sign. If some electrons leave the plasma column in the case of nonlinear response, then the wakefield carried by these electrons is always focusing for electron beams. This clarifies the choice of oblique injection as the baseline scenario of AWAKE experiment.\cite{NIMA-829-3}

\acknowledgements

This work is supported by The Russian Science Foundation, grant No.~14-50-00080. The computer simulations are made at Siberian Supercomputer Center SB RAS and Computing Center of Novosibirsk State University.

\end{document}